\pdfoutput=1
\documentclass[reprint,aip,apl, twocolumn,floatfix,superscriptaddress]{revtex4-1}
\usepackage{graphicx}
\usepackage{dcolumn}
\usepackage{bm}
\usepackage{amsmath}
\usepackage{lineno}
\usepackage[english]{babel}
\usepackage[utf8]{inputenc}
\usepackage{color}
\usepackage{siunitx}
\usepackage{amsmath}
\usepackage{textcomp, gensymb}
\usepackage{upgreek}
\usepackage{hyperref}
\hypersetup{
    colorlinks=true,
    citecolor=magenta,
    linkcolor=blue,
    filecolor=magenta,      
    urlcolor=magenta,
    }

\begin{document}

\title{Using Landau quantization to probe disorder in semiconductor heterostructures}

\author{Asser Elsayed}
\affiliation{QuTech and Kavli Institute of Nanoscience, Delft University of Technology, Lorentzweg 1, 2628 CJ Delft, Netherlands}
\author{Davide Costa}
\affiliation{QuTech and Kavli Institute of Nanoscience, Delft University of Technology, Lorentzweg 1, 2628 CJ Delft, Netherlands}
\author{Lucas E. A. Stehouwer}
\affiliation{QuTech and Kavli Institute of Nanoscience, Delft University of Technology, Lorentzweg 1, 2628 CJ Delft, Netherlands}
\author{Alberto Tosato}
\affiliation{QuTech and Kavli Institute of Nanoscience, Delft University of Technology, Lorentzweg 1, 2628 CJ Delft, Netherlands}
\author{Mario Lodari}
\affiliation{QuTech and Kavli Institute of Nanoscience, Delft University of Technology, Lorentzweg 1, 2628 CJ Delft, Netherlands}
\author{Brian Paquelet Wuetz}
\affiliation{QuTech and Kavli Institute of Nanoscience, Delft University of Technology, Lorentzweg 1, 2628 CJ Delft, Netherlands}
\author{Davide Degli Esposti}
\affiliation{QuTech and Kavli Institute of Nanoscience, Delft University of Technology, Lorentzweg 1, 2628 CJ Delft, Netherlands}
\author{Giordano Scappucci}
\affiliation{QuTech and Kavli Institute of Nanoscience, Delft University of Technology, Lorentzweg 1, 2628 CJ Delft, Netherlands}

\date{\today}

\begin{abstract}
Understanding scattering mechanisms in semiconductor heterostructures is crucial to reducing sources of disorder and ensuring high yield and uniformity in large spin qubit arrays. Disorder of the parent two-dimensional electron or hole gas is commonly
estimated by the critical, percolation-driven density associated with the metal-insulator transition.
However, a reliable estimation of the critical density within percolation theory is hindered by the need to measure conductivity with high precision at low carrier densities, where experiments are most difficult.
Here, we connect experimentally percolation density and quantum Hall plateau width, in line with an earlier heuristic intuition,  and offer an alternative method for characterizing semiconductor heterostructure disorder.
\end{abstract}

\maketitle

The ability to probe disorder from random potentials is crucial to accelerate the development cycle of semiconductor heterostructures for spin qubits in gate-defined quantum dots \cite{de2021materials, burkard2023semiconductor}. When the conductive channel in the heterostructure is separated from the semiconductor-dielectric interface by an epitaxial barrier, the typical sources of disorder affecting the electrical quality of the 2D charge-carrier gas include uniform background charged impurities within or near the channel and remote charged impurities at the semiconductor-dielectric interface \cite{scappucci_crystalline_2021}.
As a consequence of the disorder potential landscape, a metal-insulator transition occurs in the 2D charge-carrier gas \cite{abrahams2001metallic,sarma2005so,ahn_density-tuned_2023}. This transition is characterized by a critical density $n\mathrm{_c}$ that separates an effective metallic phase from an effective insulating phase and its origin has been established as a density-inhomogeneity-driven percolation transition \cite{tracy_observation_2009,das_sarma_two-dimensional_2013,das_sarma_two-dimensional_2014}
. As such, it has become common practice in the electrical characterization of heterostructures to evaluate at zero magnetic field $B$ the percolation behaviour of the density-dependent conductivity $\sigma_{\mathrm{xx}}(n) \propto (n - n_{\mathrm{c}})^{\alpha}$  \cite{last1971percolation,shklovskii1975percolation,fogelholm1980conductivity}, where $n\mathrm{_c}$ and $\alpha$ are the percolation transition density and exponent. The obtained $n\mathrm{_c}$ is considered a key indicator of disorder in the low-density regime, which is relevant for single charge occupation in quantum dots. Importantly, quantum dots with a diameter of approximately $ 1/ \sqrt{n_\mathrm{c}}$\cite{das2013two}, which indicates the average distance between impurities, could be essentially disorder-free. However, a reliable evaluation of $n\mathrm{_c}$ is challenged theoretically by the choice of the percolation exponent and the density range used for fitting, as well as experimentally by the difficulty of measuring the charge density $n$ using the Hall effect\cite{costa2024reducing, lodari_low_2021}, due to the increasing channel and contact resistance as $n\mathrm{_c}$ is approached from the high-density side of the transition.

In this Letter, we evaluate disorder in semiconductor heterostructures by demonstrating the connection between the percolation transition density and the width of the plateau of the integer quantum Hall effect, in line with earlier heuristic intuitions~\cite{efros1989metal}. Unlike the common percolation fits, we approach the critical density from the low-density side of the metal-insulator transition. Conceptually, we rely on Landau quantization to precisely tune the charge density $eB/h$ within a Landau level via magnetic fields, filling the localized states inside the disorder-induced mobility gap. 

\begin{figure}[t]
	\includegraphics[width=86mm]{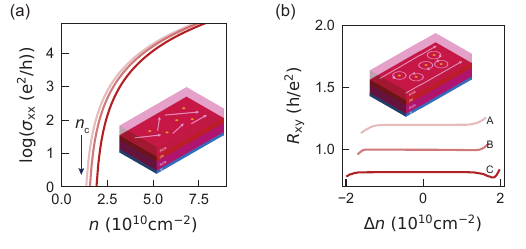}
\caption{(a) Fit to the longitudinal conductivity ($\sigma_{\mathrm{xx}}$) as a function of density ($n$) for three different strained Ge quantum well heterostructures (A, B, C) with increasing percolation density $n_\mathrm{c}$. Light red curve (heterostructure A) is from Ref.~\cite{stehouwer2023germanium}, $n_\mathrm{c} = \mathrm{1.2\times 10^{10}\ cm^{-2}}$; red curve (heterostructure B) from Ref.~\cite{costa2025buried}, $n_\mathrm{c} = \mathrm{1.4\times 10^{10}\ cm^{-2}}$; dark red curve (heterostructure C) from Ref.~\cite{lodari2022lightly}, $n_\mathrm{c} = \mathrm{1.8\times 10^{10}\ cm^{-2}}$. Inset: a schematic representation of the classically measured percolation density.
(b) For the same heterostructures, measurements of the transverse resistance ($R_\mathrm{xy}$) as a function of the carrier density deviation $\Delta n= (eB/h-n)$ from the integer filling factor condition $\nu = 1$.
Samples with lower percolation densities show narrower quantum Hall plateau.  
Measurements for heterostructures A and C are offset for clarity. Inset: a schematic representing how the cyclotron orbits are pinned around the same defects represented in the schematic in panel (a).}
\label{fig:one}
\end{figure}

Early investigations of the relationship between quantum Hall plateau width and the transport properties of two-dimensional charge-carrier gases were constrained by the use of modulation-doped structure with fixed carrier density~\cite{huckestein1995scaling}.
However, recent developments in spin qubits using gate-defined semiconductor quantum dots have led to numerous reports of percolation fits of the density-dependent conductivity in gated Hall-bars~\cite{stehouwer2023germanium, costa2025buried, lodari2022lightly, tosato_highmobility_2022, sammak_low_2019, paquelet2020multiplexed, degli2022wafer, manfra2007transport}, with quantum Hall effect measurements provided as additional side information.
In Fig.~\ref{fig:one} we consider measurements from high-mobility two-dimensional hole gases (2DHG) in Ge/SiGe heterostructures~\cite{scappucci2021germanium} to make a first qualitative connection between percolation density and quantum Hall plateaus. In Fig.~\ref{fig:one}(a), devices from 3 different heterostructures A~\cite{stehouwer2023germanium}, B~\cite{costa2025buried} and C~\cite{lodari2022lightly} of decreasing electrical quality show percolation fits of the density-dependent conductivity $\sigma_{\mathrm{xx}}(n)$ that shift to increasingly higher density, yielding larger $n\mathrm{_c}$. 
Figure~\ref{fig:one}(b) shows, for the same devices, the quantum Hall plateaus at filling factor $\nu =1$ in the transverse resistance $R_\mathrm{xy}$, measured at varying magnetic field $B$ and a density $n$ that is fixed, but different across devices. To facilitate a comparison, we plot $R_\mathrm{xy}$ as a function of $\Delta n= (eB/h-n)$, which represents the carrier density deviation from an integer-filled Landau level. Here $\Delta n=0$ corresponds to the integer filling factor condition $\nu=1$, $\Delta n>0$ measures the density of empty states within the partially filled first energy level ($\nu<1$), and $\Delta n<0$ indicates the filled states of the partially filled second energy level ($\nu>1$). A direct comparison between Fig.~\ref{fig:one}(a) and (b) establishes a clear trend: samples with lower percolation densities, and hence less disorder, show narrower plateaus. As illustrated in the insets of Fig. \ref{fig:one}, this observation underscores the role of defects in stabilizing quantum Hall effect plateaus\cite{yi2025integer}.

The connection between the widths of integer quantum Hall plateaus and the critical density of the metal–insulator transition, in the framework of percolation theory, has been approached heuristically by Efros \cite{efros1989metal}. At zero magnetic field, the transition occurs at a critical density $n_\mathrm{c}$, which reflects the percolation threshold of the disorder potential. In finite fields, the plateau widths should scale with $n_\mathrm{c}$, linking transport in the quantum Hall regime to the $B=0$ percolation transition \cite{efros1989metal}. Specifically, at low fields, multiple Landau levels contribute to screening, and the plateau width $n_\mathrm{LL}$, expressed as a density with respect to the integer filling condition, is expected to follow $n_\mathrm{LL} \sim n_c \Delta \kappa / e^2 \sqrt{C}$, where $\kappa$ is the dielectric constant, $\Delta \sim B$ is the cyclotron energy, and $C$ is the randomly distributed remote impurity density, which is mainly located at the semiconductor-dielectric interface in contemporary heterostructures used for spin qubits~\cite{scappucci_crystalline_2021}. At high fields, a condition typically met for $\nu=1$, the plateau widths become insensitive to $B$ and saturate at approximately $2n_\mathrm{c}$.

\begin{figure}
	\includegraphics[width=86mm]{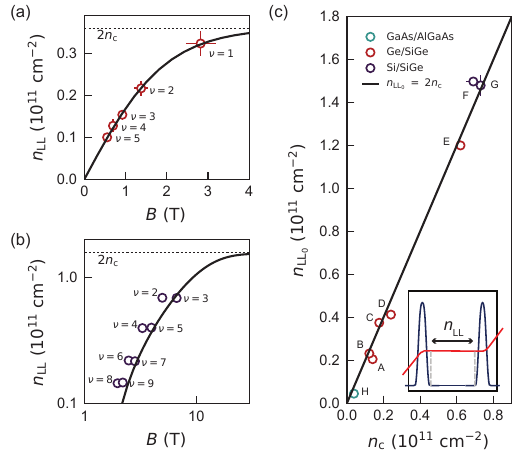}
	\caption{(a), (b) Width of the quantum Hall plateaus ($n_{\mathrm{LL}}$) at different filling factors ($\nu$) as a function of magnetic field ($B$), shown for holes in Ge/SiGe (red, heterostructure C from Ref.~\cite{lodari2022lightly}) and electrons in Si/SiGe heterostructures (purple, heterostructure G from Ref.~\cite{degli2022wafer}), respectively. The fit to Eq.\ref{eq:Plateau} (black lines) gives an asymptotic width of the quantum Hall plateaus ($n_{\mathrm{LL_{0}}}$), not far from the experimental values for $\nu=1$. 
    The fit for electron in silicon considers only the odd filling factors, which correspond to valley split levels. 
    The dashed lines are placed at twice the percolation density $n_{\mathrm{c}}$ measured for each sample.
    (c) Asymptotic width of the QHE plateau $n_{\mathrm{LL_{0}}}$ as a function of percolation density $n_{\mathrm{c}}$ for Ge/SiGe heterostructures (red), GaAs/AlGaAs heterostructures (green) and Si/SiGe heterostructures (purple). Heterostructure A is from Ref.~\cite{stehouwer2023germanium}; heterostructure B is from Ref.~\cite{costa2025buried}; heterostructure C is from Ref.~\cite{lodari2022lightly}; heterostructure D is from Ref.~\cite{tosato_highmobility_2022}; heterostructure E is from Ref.~\cite{sammak_low_2019}; heterostructure F is from Ref.~\cite{paquelet2020multiplexed}; heterostructure G is from Ref.~\cite{degli2022wafer} and heterostructure H is from Ref.~\cite{manfra2007transport}. The black solid line corresponds to $n_{\mathrm{LL_0}} = 2n_{\mathrm{c}}$ as suggested by Efros~\cite{efros1989metal}.}
\label{fig:two}
\end{figure}

We investigate this correlation systematically by using data from the 2DHG in Ge/SiGe in Ref.~\cite{lodari2022lightly} measured at a temperature below 100~mK.  We extract the widths of quantum Hall plateaus $n_{\mathrm{LL}}$ at increasing filling factors and plot in them Fig.~\ref{fig:two} against the magnetic field value corresponding to integer filling factor. Here, the width refers specifically to the region where the longitudinal resistivity ($\rho_{\mathrm{xx}}$) is zero.
At low fields, $n_\mathrm{LL}$ increases approximately linearly with $B$, while at high fields it approaches a constant. To capture both regimes, we fit the data to the model:

\begin{equation}
    n_{\mathrm{LL}} = n_{\mathrm{LL_0}}\tanh\left(\frac{n_{\mathrm{LL_0}}\Delta\kappa}{2e^2\sqrt{C}}\right)
    \label{eq:Plateau}
\end{equation}

which interpolates smoothly between the low- and high-field limits. From this fit, we extract a characteristic values $n_{\mathrm{LL_0}}$ of $\mathrm{3.6\times 10^{10}\ cm^{-2}}$ and a remote impurity density $C$ of $\mathrm{3.2\times 10^{11}\ cm^{-2}}$, which is reasonable considering prior studies on similar heterostructures field effect transistors~\cite{massai_impact_2024}. Our analysis shows that $n_{\mathrm{LL_0}}$ is approximately twice the critical density $n_\mathrm{c}$ = $\mathrm{1.76\times 10^{10}\ cm^{-2}}$ (black dotted line) obtained independently from percolation fits in Ref~\cite{lodari2022lightly}, in line with the suggestion of Efros~\cite{efros1989metal}.

Similarly, we preform the same analysis on data from a 2D electron gas in Si/SiGe from Ref.\cite{degli2022wafer}, also measured at a temperature below 100~mK, and show the results in Fig. \ref{fig:two}b. We observe that the even and odd filling factors have similar plateau widths, suggesting additional effects beyond the simplistic theoretical approach of Efros\cite{efros1989metal}, possibly related due to the additional valley degree of freedom  which complicates the Landau level energy ladder in silicon\cite{wuetz2020effect, lodari2022valley}. For this reason, we restrict the fit to Eq. \ref{eq:Plateau} to  the valley-resolved odd filling factors 3, 5, 7, 9. As for holes in germanium, also for electrons in silicon the data aligns with the model, and we find that the fitted $n_{\mathrm{LL_0}}$ of $\mathrm{1.56\times 10^{11}\ cm^{-2}}$ is approximately twice the $n_\mathrm{c}$ value of $\mathrm{0.78\times 10^{11}\ cm^{-2}}$ ((black dotted line) reported in Ref.~\cite{degli2022wafer}.

In Fig. \ref{fig:two}c we extend the same analysis to a range of heterostructures, including 2DEGs in GaAs/AlGaAs \cite{manfra2007transport} and Si/SiGe \cite{paquelet2020multiplexed, degli2022wafer}, as well as 2DHGs in Ge/SiGe \cite{lodari2022lightly, sammak_low_2019, stehouwer2023germanium, tosato_highmobility_2022, costa2025buried}, for which percolation densities and quantum Hall effect measurements have been previously reported at milliKelvin temperatures.
We observe that the extracted $n_{\mathrm{LL_0}}$ is consistently about twice the reported $n_\mathrm{c}$ across all systems~\footnote{A linear fit of $n_{\mathrm{LL_0}}$ against $n_\mathrm{c}$ yields a slope of 2.12(7) and an intercept of -0.04(4).}, overlapping with the theoretical prediction $n_{\mathrm{LL_0}} = 2n_{\mathrm{c}}$ (black line), spanning about an order of magnitude in percolation density. This result confirms the correlation between percolation density measured at low charge density and zero magnetic field, and the disorder effects observed in the QHE at overall high charge density $p$ and strong magnetic fields. 

\begin{figure}[t]
	\includegraphics[width=86mm]{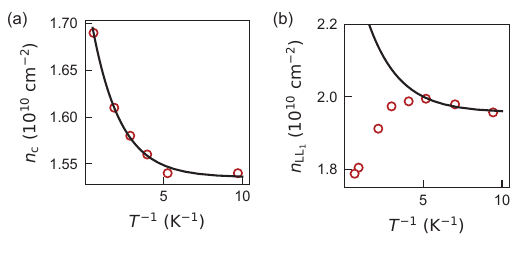}
	\caption{(a) Percolation density $n_\mathrm{c}$ as a function of inverse temperature $1/T$.  The data points are obtained from  percolation fits, with conductivity measurements performed on a device from heterostructure A, Ref. \cite{stehouwer2023germanium}. The solid line is a fit to Eq.~\ref{eq:temp}. (b) Width of the quantum Hall plateau for the first filling factor $n_\mathrm{LL_{1}}$ as a function of inverse temperature (same sample as in panel (a)). The solid line is an activation energy fit considering the three coldest temperatures. At higher temperature, Landau level broadening smears out the quantum Hall plateaus.} 
\label{fig:three}
\end{figure}

Having established the correlation between the classical and quantum transport regimes in quantum wells, we give further insights into the temperature dependence on this relationship. As the temperature increases, the dielectric function changes and the screening length decreases, effectively activating additional defects~\cite{kruithof1991temperature}. These defects influence both the scattering properties of the two-dimensional system and the percolation density~\cite{tracy_observation_2009,kim2017annealing}. In Fig.~\ref{fig:three}a, we present $n_{\mathrm{c}}$ as a function of temperature. The data, from 2DHG in Ge/SiGe in Ref~\cite{stehouwer2023germanium}, show a decrease in percolation density with decreasing temperature, eventually reaching saturation at low temperatures. The results are fitted to an activation energy model in the form of:

\begin{equation}
    n_{\mathrm{c}} = n_{\mathrm{0}} + A\exp{\left(\frac{-b}{T}\right)}
    \label{eq:temp}
\end{equation}
from the fit we find the activation energy to be roughly 48 $\upmu\mathrm{eV}$. 

The Landau levels in the QHE are broadened by both collisional and thermal effects\cite{isihara1986density, fang1968effects, paalanen1982quantized}. As the temperature increases, the thermal energy $k_BT$ becomes comparable to the energy gap between Landau levels, leading to thermal broadening. In this regime, electrons and holes can occupy higher energy states, diminishing the discrete nature of the Landau levels. This increases the likelihood of carriers scattering between levels, which disrupts the quantized Hall conductance.
In Fig. \ref{fig:three}b, we observe the behavior of the $\nu=1$ plateau width ($n_{\mathrm{LL_1}}$) as a function of temperature. Initially, the plateau width increases with temperature due to enhanced collisional broadening. However, as thermal broadening becomes dominant, the plateau width begins to decrease once $k_BT$ approaches the Landau level energy gap. This restricts the determination of percolation density to low temperatures, where collisional broadening outweighs thermal broadening.

In summary, we consider a variety of two-dimensional electron and hole gases and demonstrate the connection between percolation density extracted from the density-dependent conductivity at zero magnetic field and the Landau level plateau widths at high magnetic fields, in line with earlier heuristic intuitions. This method offers an alternative approach to characterizing disorder in semiconductor heterostructures, critical for spin qubit development. However, this method requires magnetic fields high enough to resolve the first filling factor and low temperatures where Landau levels are collisional- rather than thermal-broadened. 

\bigskip

We acknowledge support by the European Union through the IGNITE project with grant agreement No. 101069515 and the QLSI project with grant agreement No. 951852.
This work was supported by the Netherlands Organisation for Scientific Research (NWO/OCW), via the Open Competition Domain Science - M program. A.T. acknowledges the research programme Materials for the Quantum Age (QuMat) for financial support. This programme (registration no. 024.005.006) is part of the Gravitation programme financed by the Dutch Ministry of Education, Culture and Science (OCW). This research was sponsored in part by The Netherlands Ministry of Defence under Awards No. QuBits R23/009. The views, conclusions, and recommendations contained in this document are those of the authors and are not necessarily endorsed nor should they be interpreted as representing the official policies, either expressed or implied, of The Netherlands Ministry of Defence. The Netherlands Ministry of Defence is authorized to reproduce and distribute reprints for Government purposes notwithstanding any copyright notation herein.

\section*{Author Declarations}
 G.S. is founding advisor of Groove Quantum BV and declares equity interests.

\section*{Data availability}
The data sets supporting the findings of this study are
openly available at the Zenodo repository~\cite{repo2025}

\end{document}